\documentclass[11pt,nofootinbib,showpacs]{revtex4}
\usepackage{amsfonts,amssymb,amsmath,graphicx}
\def\dac{\displaystyle\frac}

\def\[{\left[}
\def\]{\right]}
\def\({\left(}
\def\){\right)}

\def\gammaterm{\gamma _{\left( \mathbf{D}\right)}+\dot b(t)^2}
\def\bb{b(t)}
\def\db{\dot b(t)}
\def\ddb{\ddot b(t)}
\def\hh{H(t)}
\def\dh{\dot H(t)}

\newcommand{\const}{\mathop{\rm const}\nolimits}
\begin{document}

\baselineskip7mm

\title{Friedmann dynamics recovered from compactified Einstein-Gauss-Bonnet cosmology}

\author{Fabrizio Canfora}
\affiliation{Centro de Estudios Cientificos (CECs), Casilla 1469 Valdivia, Chile}
\author{Alex Giacomini}
\affiliation{Instituto de Ciencias F\'isicas y Matem\'aticas, Universidad Austral de Chile, Valdivia, Chile}
\author{Sergey A. Pavluchenko}
\affiliation{Programa de P\'os-Gradua\c{c}\~ao em F\'isica, Universidade Federal do Maranh\~ao (UFMA), 65085-580, S\~ao Lu\'is, Maranh\~ao, Brazil}
\author{Alexey Toporensky}
\affiliation{Sternberg Astronomical Institute, Moscow State University, Moscow 119991 Russia}
\affiliation{Kazan Federal University, Kazan 420008 Russia}

\begin{abstract}
In this paper cosmological dynamics in Einstein-Gauss-Bonnet gravity with a perfect fluid source in arbitrary dimension is studied. A systematic analysis is performed for the case that the
theory does not admit maximally symmetric solutions. Considering  two independent scale factors, namely one for the three dimensional space and one for the extra dimensional space, is found
that a regime exists where the two scale factors  tend to a constant value via damped oscillations for not too negative pressure of the fluid,
so that asymptotically the evolution of the $(3+1)$-dimensional Friedmann model with perfect
fluid is recovered.
At last, it is worth emphasizing that the present numerical results strongly support a 't Hooft-like interpretation of the parameter $1/D$ (where $D$ is the number of extra dimensions) as a 
small expansion parameter in very much the same way as it happens in the large $N$ expansion of gauge theories with $1/N$. Indeed, the dependence on $D$ of many of the relevant physical 
quantities computed here manifests a clear WKB-like pattern, as expected on the basis of large $N$ arguments.
\end{abstract}







\maketitle

\section{Introduction}
One of the outstanding conceptual features of General Relativity (GR) is that the geometry of space-time itself is a dynamical object. This opens the possibility that there may exist more than
three space dimensions which are not observable because they may be compactified to a very small scale. This idea was implemented the first time by Kaluza~\cite{KK1} and Klein~\cite{KK2, KK3}
assuming one extra space dimensions in order to attempt to unify gravity with electromagnetism. This idea can be extended to include non-Abelian gauge fields by introducing more extra
dimensions. Moreover the existence of extra dimensions is also predicted by String Theory. The low energy sector of some String theories  is not described by General Relativity but
by a generalization of it known in literature as Einstein-Gauss-Bonnet gravity (see e.g. \cite{GastGarr}). The action of this theory is the sum of three terms two of which are the familiar
$\Lambda$-term and the Einstein-Hilbert term whereas the third term is the Gauss-Bonnet term which is quadratic in the curvature and reads
$I_{GB}=\int \sqrt{-g} (R^2-4R_{\mu \nu}R^{\mu \nu}+R_{\alpha \beta  \mu \nu}R^{\alpha \beta \mu \nu})$. In four dimension the Gauss-Bonnet term is topological and does not affect the equations
of motion. In dimension higher than four however this term leads to a non-trivial contribution to equation of motion. As this non-trivial contribution is still of second order in the derivatives
of the metric, Einstein-Gauss-Bonnet gravity can be considered as a natural extension of General Relativity to higher dimensions as its action is constructed according to the same principles as
the Einstein Hilbert action in four dimensions. Einstein-Gauss-Bonnet gravity is actually a particular case of a more general gravity theory known as Lovelock gravity \cite{Lovelock}. In the
same way as the Gauss-Bonnet term, which is topological in four dimensions but gives a non-trivial contribution to the equations of motion in higher dimension in every odd dimension it is
possible to add a new higher power term in the curvature which is topological in the lower even dimension (this means for example that in seven dimensions the most generic Lovelock theory is
sum of four terms namely the three terms of Einstein-Gauss-Bonnet gravity and a new term which is cubic the curvature). All these higher power terms again lead to a non-trivial contribution to
the equations of motion which is of second order in the derivatives.

Even if Lovelock gravity is constructed according to the same principles as General Relativity it has some features which are absent in General Relativity.
For example in the first order formalism the equation of motion do not imply the vanishing of torsion \cite{TZ-CQG} which therefore becomes a new propagating degree of freedom. Exact solution
with non-trivial vacuum torsion have been found in \cite{CGT07, CGW07, CG, CG2, ACGO}. In this paper however we will consider the zero torsion sector.

Perhaps one of the most new feature of Lovelock gravity is that the ``Lambda term'' in the action is not directly related to the cosmological constant in the sense that it does not
measure the curvature of the maximally symmetric solution. Actually for a $N$th order Lovelock theory there can exist up to to $N$ different maximally symmetric solutions where the curvature
radii
are function of all Lovelock couplings. A very peculiar case happens when the highest Lovelock term in the action is of even power in the curvature as there may exist no maximally symmetric
solution at all so that the vacuum state must be necessarily a less symmetric space-time. This situation can be interpreted as a case of ``geometric frustration'' (see e.g.~\cite{CGP1, CGP2}).

In order to recover physics in four space-time dimensions it is necessary to check if the equations of motion of Lovelock gravity admit compactified solutions known in literature as
``spontaneous compactification''.   Exact static solutions where the metric is a cross product of a (3+1)-dimensional manifold and a constant curvature
``inner space'',  were discussed for the first time in~\cite{add_1}, but with (3+1)-dimensional manifold being actually Minkowski (the generalization for a constant
curvature Lorentzian manifold was done in~\cite{Deruelle2}). It was shown in \cite{CGTW09} that in order to recover four dimensional General Relativity  with arbitrarily small cosmological
constant and with arbitrarily small static extra dimensions it is necessary to have a Lovelock theory which includes also at least a cubic term so that the minimal space-time dimension where
such a spontaneous compactification can happen is seven.

In the context of cosmology it is of course of fundamental interest to consider a spontaneous compactification where the four dimensional part is given by a Friedmann-Robertson-Walker metric.
In this case it is then completely natural to consider also the size of the extra dimensions as time dependent rather then static. Indeed in
\cite{add_4} it was explicitly shown  that in order to have a more realistic model one needs to consider the dynamical evolution of the extra dimensional scale factor as well.
In~\cite{Deruelle2} the equations of motion for compactification with both time dependent scale factors were written for arbitrary Lovelock order in the special case that both factors are flat.
The results of~\cite{Deruelle2} were reanalyzed for the special case of 10 space-time dimensions in~\cite{add_10}.
In~\cite{add_8} the existence of dynamical compactification solutions was studied with the use of Hamiltonian formalism.
More recently efforts in finding spontaneous  compactifications have been done in  \cite{add13} where
the dynamical compactification of (5+1) Einstein-Gauss-Bonnet (EGB) model was considered, in \cite{MO04, MO14} with different metric {\it ansatz} for scale factors
corresponding to (3+1)- and extra dimensional parts, and in \cite{CGP1, CGP2} where general (e.g. without any {\it ansatz}) scale factors and curved manifolds were considered. Also, apart from
cosmology recent analysis focuses on
properties of black holes in Gauss-Bonnet~\cite{addn_1, addn_2} and Lovelock~\cite{addn_3, addn_4} gravities, features of gravitational collapse in these
theories~\cite{addn_5, addn_6, addn_7}, general features of spherical-symmetric solutions~\cite{addn_8} and many others.

The most common ansatz used to find exact solutions for the functional form of the scale factor were exponentials or power laws.
Exact solutions with exponential functions  for both  the (3+1)- and extra dimensional scale factors were studied the first time in  \cite{Is86}  and exponentially increasing (3+1)-dimensional
scale factor and exponentially shrinking extra dimensional scale factor were described.
Power-law solutions have been analyzed  in \cite{Deruelle1, Deruelle2} and more  recently~\cite{mpla09, prd09, Ivashchuk, prd10, grg10} so that there is an  almost complete description
(see also~\cite{PT} for useful comments regarding physical branches of the solutions).
More recently solutions with exponential scale factors~\cite{KPT} have been studied in detail namely models with both variable~\cite{CPT1} and constant~\cite{CST2} volume developing a general
scheme for
constructing solutions in EGB; recently~\cite{CPT3} this scheme
was generalized for general Lovelock gravity of any order and in any dimensions as well. Also, the stability of the solutions was addressed in~\cite{my15}, where it was
demonstrated that only a handful of the solutions could be called ``stable'' while the remaining are either unstable or have neutral/marginal stability and so additional investigation is
required.

In order to find all possible regimes of Einstein-Gauss-Bonnet cosmology it is necessary to go beyond an exponential or power law ansatz and keep the functional form of the scale factor generic.
In this case the equations of motion are too difficult to be solved analytically and so a numerical analysis must be performed. In \cite{CGP1} it was found that there exist a phenomenologically
sensible regime  in the case that the curvature of the extra dimensions is negative and the Einstein-Gauss-Bonnet theory does not admit a maximally symmetric solution. In this case the
three dimensional Hubble parameter and the extra dimensional scale factor tend to  constant values asymptotically. Due to the fact that the theory does not admit maximally symmetric solution
the compactification was interpreted as geometric frustration. The reduced symmetry vacua of a Lovelock theory without maximally symmetric solution have been explored in~\cite{castor_2015}.
In \cite{CGP2} a detailed analysis of the cosmological dynamics of Einstein-Gauss-Bonnet gravity with generic couplings was performed. In most cases it turned out that there  exist non-physical
features like  a finite time  future singularity or isotropization. The only situation where a realistic dynamics could be achieved turned out to be the case when the theory does not admit
maximally symmetric solutions (and with negative curvature of the extra dimensions).

A further benefit to use two different scale factors for the three ``macroscopic'' dimensions and the $D$ extra-dimensions is that it allows to disclose a clear ``large $N$ pattern'' in the
dependence of relevant physical observables on the number of extra-dimensions. In gauge theories the large $N$ expansions introduced by 't Hooft in~\cite{thooft}(and generalized by Veneziano
in~\cite{veneziano} to take into account the presence of matter fields in the fundamental) provides with a very clever small parameter suitable to generate an alternative perturbative expansion
(namely $1/N$) which does not necessarily require the theory to be weakly coupled. A crucial step to achieve this goal is to separate in a very clear way the dependence on internal indices from
the dependence on ``space-time'' indices of all relevant quantities (for two nice reviews see~\cite{MAKMAN}). Such large $N$ expansion is very closely related to the semi-classical or WKB
expansion and is a very powerful non-perturbative tool in field theory. In (3+1) dimensional General Relativity the situation is complicated by the fact that it is not easy to distinguish
``internal'' from ``space-time'' indices in the second order formalism (in the first order formalism the results in~\cite{canfora2005} suggest that the 't Hooft expansion in gravity in (3+1)
dimensions has many features absent in the Yang-Mills case). However, in the case of Kaluza-Klein theories, there is a parameter which is clearly analogous to the 't Hooft parameter $N$:
the number of extra dimensions~\cite{CGZ}. The results of the present paper clearly suggest that the same is true for EGB theory. For this reason, the field equations have been written in order
to display explicitly the dependence on $D$. The present analysis shows that the dependence of many relevant quantities on the number of extra dimensions exhibit a WKB-like pattern.

Of course in  cosmology it is also crucial to study the effect of a matter source on the cosmological dynamics.
For this reason the purpose of this paper is to explore the effect of a perfect fluid on Einstein-Gauss-Bonnet cosmology in the geometric frustration regime. Especially the emergence of new
subregimes is studied. It is found that there exist a new subregime where both scale factors tend to constant values by performing damped oscillations  so that the Friedmann dynamics can be
recovered asymptotically.  It is also worth to point out that the dependence of the relevant physical quantities on the number of the extra-dimensions reminds very closely what one would
expect from a large $N$ expansion in gauge theories in which the number of extra-dimensions plays the role of $N$ in the 't Hooft expansion. In particular, the present analysis discloses a
clear eikonal-like pattern.

The structure of the paper will be the following: in section 2,  some analytic considerations are given of why it is necessary to consider the geometric frustration regime. In section 3, the
subregime which can recover $(3+1)$-dimensional Friedmann model is introduced. In section 4 matter is added and the numerical analysis is performed. Section 5 is dedicated to the conclusions.

\section{Equations of motion}

Lovelock gravity~\cite{Lovelock} has the following structure: its Lagrangian is constructed from terms

\begin{equation}
L_n = \frac{1}{2^n}\delta^{i_1 i_2 \dots i_{2n}}_{j_1 j_2 \dots
j_{2n}} R^{j_1 j_2}_{i_1 i_2}
 \dots R^{j_{2n-1} j_{2n}}_{i_{2n-1} i_{2n}}, \label{lov_lagr}
\end{equation}

\noindent where $\delta^{i_1 i_2 \dots i_{2n}}_{j_1 j_2 \dots
j_{2n}}$ is the generalized Kronecker delta of the order $2n$.
One can verify that $L_n$ is Euler invariant in $D < 2n$ spatial dimensions and so it would not give nontrivial contribution into the equations of motion. So that the
Lagrangian density for any given $D$ spatial dimensions is sum of all Lovelock invariants (\ref{lov_lagr}) upto $n=\[\dac{D}{2}\]$ which give nontrivial contributions
into equations of motion:

\begin{equation}
{\cal L}= \sqrt{-g} \sum_n c_n L_n, \label{lagr}
\end{equation}

\noindent where $g$ is the determinant of metric tensor,
$c_n$ is a coupling constant of the order of Planck length in $2n$
dimensions and summation over all $n$ in consideration is assumed.

The {\it ansatz} for the metric is
\begin{equation}
ds^{2}=-dt^{2}+a(t)^{2}d\Sigma _{(3)}^{2}+b(t)^{2}d\Sigma _{(\mathbf{D}%
)}^{2}\ ,  \label{Ansatz-metric}
\end{equation}%

\noindent where $d\Sigma _{(3)}^{2}$ and $d\Sigma _{(\mathbf{D})}^{2}$ stand for the
metric of two constant curvature manifolds $\Sigma _{(3)}$ and $\Sigma_{(\mathbf{D})}$\footnote{We consider {\it ansatz} for space-time in form of a warped product
\mbox{$M_4\times b(t)M_D$}, where $M_4$ is a Friedmann-Robertson-Walker manifold with scale factor $a(t)$ whereas
$M_D$ is a $D$-dimensional Euclidean compact and constant curvature manifold with scale factor $b(t)$.}. It is worth to point out that even a negative constant curvature space can be
compactified by making the quotient of the space by a
freely acting discrete subgroup of $O(D,1)$ \cite{wolf}.

The complete derivation of the equations of motion could be found in our previous papers, dedicated to the description of the particular regime which appears in this model~\cite{CGP1, CGP2}.
For simplicity we consider the case with $\gamma _{\left( 3\right)} = 0$ (zero spatial curvature for ``our'' (3+1)-dimensional world; nonzero $\gamma _{\left( 3\right)}$ does not affect the
presence of the dynamical compactification regime, as discussed in~\cite{CGP1, CGP2}).
For the moment the non-zero curvature for extra dimensions can be normalized
as $\gamma _{\left( \mathbf{D}\right)} = \pm 1$. Since there is no curvature
term for $a(t)$, it is useful to rewrite the equations of motion in terms of the Hubble parameter $H(t) = \dot a(t)/a(t)$; the equations will take a form


\small
\begin{equation}
\begin{array}{l}
\dac{(D+3)(D+2)(D+1)}{6}c_0 + \dac{(D+1)D(D-1)}{6}c_1 \[ \dac{\gammaterm}{\bb^2} + \dac{6\hh\db}{\bb(D-1)} + \dac{6\hh^2}{D(D-1)}  \] + \\ \\ +
\dac{(D-1)(D-2)(D-3)}{6}c_2 \[ \dac{(\gammaterm)^2}{\bb^4} + \dac{12\hh^2(\gammaterm)}{\bb^2(D-2)(D-3)} + \dac{24\hh^2\db^2}{\bb^2(D-2)(D-3)} + \right. \\ \\ \left. +
\dac{12(\gammaterm)\hh\db}{\bb^3(D-3)} + \dac{24\hh^3\db}{\bb(D-1)(D-2)(D-3)} \]=0,
\end{array}\label{eq_my_1}
\end{equation}

\begin{equation}
\begin{array}{l}
\dac{(D+3)(D+2)(D+1)}{6}c_0 + \dac{(D+1)D(D-1)}{6}c_1 \[ \dac{\gammaterm}{\bb^2} + \dac{4(\dh+\hh^2)}{D(D-1)} + \dac{2\ddb}{\bb(D-1)} + \right. \\ \\ \left. +\dac{2\hh^2}{D(D-1)} +
\dac{4\hh\db}{\bb(D-1)}  \] + \dac{(D-1)(D-2)(D-3)}{6}c_2 \[ \dac{(\gammaterm)^2}{\bb^4} + \right. \\ \\ \left. + \dac{16(\dh+\hh^2)\hh\db}{\bb(D-1)(D-2)(D-3)} +
\dac{8(\gammaterm)\hh\db}{\bb^3(D-3)} + \dac{8(\dh+\hh^2)(\gammaterm)}{\bb^2(D-2)(D-3)} + \right. \\ \\ \left. + \dac{8\hh^2\ddb}{\bb(D-1)(D-2)(D-3)} +
\dac{16\ddb\hh\db}{\bb^2(D-2)(D-3)} + \dac{4\ddb(\gammaterm)}{\bb^3(D-3)} + \right. \\ \\ \left. + \dac{4\hh^2 (\gammaterm)}{\bb^2(D-2)(D-3)} + \dac{8\hh^2\db^2}{\bb^2(D-2)(D-3)}  \]=0,
\end{array}\label{eq_my_2}
\end{equation}

\begin{equation}
\begin{array}{l}
\dac{D(D+3)(D+2)(D+1)}{6}c_0 + \dac{(D-2)(D+1)D(D-1)}{6}c_1 \[ \dac{\gammaterm}{\bb^2} + \dac{6(\dh+\hh^2)}{(D-2)(D-1)} + \right. \\ \\ \left. \dac{2\ddb}{\bb(D-2)} +
\dac{6\hh^2}{(D-2)(D-1)} +
\dac{6\hh\db}{\bb(D-2)}  \] + \dac{(D-1)(D-2)(D-3)}{6}c_2 \times \\ \\ \times \[ \dac{(\gammaterm)^2}{\bb^4}(D-4) + \dac{48(\dh+\hh^2)\hh\db}{\bb(D-2)(D-3)} +
\dac{12(\gammaterm)\hh\db}{\bb^3} + \right. \\ \\ \left. + \dac{24\hh^2\db^2}{\bb^2(D-3)} + \dac{12(\dh+\hh^2)(\gammaterm)}{\bb^2(D-3)} + \dac{24\hh^2\ddb}{\bb(D-2)(D-3)} +
\dac{24\ddb\hh\db}{\bb^2(D-3)} + \right. \\ \\ \left. + \dac{4\ddb(\gammaterm)}{\bb^3} + \dac{12\hh^2 (\gammaterm)}{\bb^2(D-3)} + \dac{24\hh^3\db}{\bb(D-2)(D-3)} +
\dac{24(\dh+\hh^2)\hh^2}{(D-1)(D-2)(D-3)} \]=0,
\end{array}\label{eq_my_3}
\end{equation}
\normalsize

\noindent as $\mathcal{E}_{0}=0$ equation (\ref{eq_my_1}), $\mathcal{E}_{i}=0$ (\ref{eq_my_2}), and  $\mathcal{E}_{a}=0$ (\ref{eq_my_3}). In~\cite{CGP1, CGP2} we described a regime which
naturally appear if a choice for coupling constants forbids existence of maximally-symmetric solutions. A maximally symmetric space-time has  curvature two form given by

\begin{equation}
R^{AB}=\Lambda_{D+4}e^{A}e^{B}\label{constantcurv},
\end{equation}

\noindent which inserted in the equations of motion gives a quadratic equation for $\Lambda_{D+4}$:

\begin{equation}
(c_2\Lambda_{D+4}^2 + c_1\Lambda_{D+4}+c_0)e^{A_1}\ldots e^{A_{D+4}}=0\label{polynomial},
\end{equation}

\noindent which admits as solutions

\begin{equation}
\Lambda_{eff}\equiv\Lambda_{D+4}= \frac{-c_1\pm \sqrt{(c_1)^2-4c_2c_0}}{2c_2}\label{Leff}.
\end{equation}

\noindent We can clearly see that if radicand in (\ref{Leff}) is negative, then effective $(4+D)$-dimensional cosmological constant is imaginary and there are no maximally-symmetric solutions
-- this is one prerequisites for the existence of our regime. The other requirement is that the spatial curvature of the extra dimensions should be negative.

\section{Vacuum regime}

The importance of the non-existence of a maximally symmetric solution for the cosmological dynamics in Einstein-Gauss-Bonnet gravity was found numerically in~\cite{CGP1, CGP2}. However it is
possible to make some analytic considerations of why this regime is relevant. Indeed it is interesting to explore the situation where the two scale factors are constant. Substituting  $ \dot H = H \equiv 0$
and $b(t) \equiv b_0$ as well as usual requirements for our regime ($\gamma_{(3)} = 0$, $\gamma_{(\mathbf{D})} = -1$, $\delta_1 = 0$) into
(\ref{eq_my_1})--(\ref{eq_my_3}), we get

\begin{equation}
\begin{array}{l}
(D+3)(D+2)(D+1) c_0 - \dac{(D+1)D(D-1)c_1}{b_0^2} + \dac{(D-1)(D-2)(D-3)c_2}{b_0^4} = 0, \\ \\
\dac{D(D+3)(D+2)(D+1) c_0}{D-4} - \dac{(D+1)D(D-1)(D-2)c_1}{b_0^2(D-4)} + \dac{(D-1)(D-2)(D-3)c_2}{b_0^4} = 0.
\end{array} \label{sys1}
\end{equation}

It should be noted that despite the Gauss-Bonnet contribution is dynamically important starting from
four spatial dimensions (i.e. $D=1$ in our notations), the regime under consideration requires bigger
number of dimensions. Indeed, we can see from Eqs. (\ref{eq_my_1})--(\ref{eq_my_3})
 that for $D<4$ all terms in equations of motion originating from the GB
term in the Lagrangian vanish if $H=0$, $b=const$. This means that the equations of motion reduce to those of GR
where this regime is absent (it is easy to see that GR equations have no solutions with $H=0$ and $b=const$).
As a result, the minimum possible value of compactified space is $D=4$, so as full space-time is at least seven dimensional.

By eliminating last terms from both equations in (\ref{sys1}) we can find $b_0$

\begin{equation}
\begin{array}{l}
b_0^2 = \dac{D(D-1)}{2(D+3)(D+2)} \dac{c_1}{c_0}
\end{array} \label{b0}
\end{equation}

\noindent and then by constitution it back to (\ref{sys1}) we can find a relation between couplings:

\begin{equation}
\begin{array}{l}
D^2 (D^2-1) c_1^2 = 4 c_0 c_2 (D^2-9) (D^2 - 4).
\end{array} \label{coup}
\end{equation}

From (\ref{b0}) we can clearly see that $c_1$ and $c_0$ should have same sign while from (\ref{coup}) -- that $c_0$ and $c_2$ should have the same property, so that all couplings should have
the same sign for this regime to occur. Additionally, if we substitute (\ref{coup}) into the equation for effective $(4+D)$ Lambda-term we can verify that its discriminant is always
negative:

\begin{equation}
\begin{array}{l}
\mathcal{D} = c_1^2 - 4c_0 c_2 = c_1^2 \( 1 - \dac{D^2 (D^2 - 1)}{(D^2 - 9)(D^2 - 4)} \) = - \dac{12c_1^2 (D^2 - 3)}{(D^2 - 9)(D^2 - 4)} < 0;
\end{array} \label{dsc1}
\end{equation}

\noindent so that this regime always occurs ``inside'' our compactification scheme. The condition (\ref{coup}) is therefore a subregime of the regime where no maximally symmetric vacuum exist.
This is of special interest in the case that for asymptotic time one wants to recover the Friedmann dynamics of four dimensions. It is worth to emphasize that, at leading order in the large $D$
expansion, the discriminant $\mathcal{D}$ vanishes so that we would be on the boundary of the region allowing our compactification scheme. However, the sub-leading large $D$ corrections make $\mathcal{D}$ more and more
negative. Thus the large $D$ expansion clearly favors our compactification scheme.

It is worth to stress here that in order to recover GR asymptotic from the constraint equation (4) it is enough to set
 the three dimensional Hubble parameter to  constant rather than the scale factor namely if we substitute $H(t) = H \equiv \const$, $b(t) = b \equiv \const$ into constraint equation
 (4) , it takes a Friedmann-like form:

\begin{equation}
\begin{array}{l}
G_{N, eff} H^2 = \Lambda_{4D, eff}.
\end{array} \label{friedmann1}
\end{equation}

This regime have been studied numerically in~\cite{CGP1, CGP2}, and it leads to monotonic behavior of both Hubble parameters when $\Lambda$ is
of the order of unity in natural units. On the contrary, here we will see that very small $H$ in natural units (which is necessary to
describe the Universe we live in) is reached through an oscillatory regime.
Our numerical studies confirm that (\ref{b0}) is holding and $b(t) \to b_0$ in oscillatory regime (see Fig.~\ref{fig1matter}(d)--(f)).

At this point it sounds useful to address linear stability of this solution. We perturb full system of equations of motion (\ref{eq_my_1})--(\ref{eq_my_3}) with small
perturbations $H(t) \to H(t) + \delta H$, $b(t) \to b(t) + \delta b$ and perturbations equation take the form

\begin{equation}
\begin{array}{l}
a_1 \delta \dot H + a_2 \delta H + a_3 \delta \ddot b + a_4 \delta \dot b + a_5 \delta b = 0, \\ b_1 \delta \dot H + b_2 \delta H + b_3 \delta \ddot b + b_4 \delta \dot b + b_5 \delta b = 0, \\
d_1 \delta H + d_2 \delta \dot b + d_3 \delta b = 0,
\end{array} \label{pert1}
\end{equation}

\noindent where first two equations correspond to dynamical equations and the last -- to constraint equation. Generally all the coefficients $a_i$, $b_i$ and $d_i$ are nonzero, but as we seek
for perturbations around our special solution, we apply $\dot H = H \equiv 0$, $\ddot b = \dot b \equiv 0$, as well as keep (\ref{b0}) and (\ref{coup}) in mind, so

\begin{equation}
\begin{array}{l}
a_2 =a_4=a_5=b_2=b_3=b_4=d_i \equiv 0, \\
a_1 = \dac{4}{3} \dac{(D+1)(2D-3)c_1}{(D-2)(D-3)},~~ a_3 = \dac{2}{3} \dac{D(D+1)\sqrt{2}c_1}{(D-3)\sqrt{\frac{D(D-1)c_1}{(D+3)(D+2)c_0}}}, \\ \\
b_1 = - \dac{2D(D+1)c_1}{(D-3)(D-4)},~ b_5 = \dac{4}{3} \dac{(D+1)(D+2)(D+3)c_0\sqrt{2}}{(D-4)\sqrt{\frac{D(D-1)c_1}{(D+3)(D+2)c_0}}}.
\end{array} \label{pert2}
\end{equation}

Now if we substitute (\ref{pert2}) into (\ref{pert1}) and solve it, the solution takes harmonic form and its frequency $f$ is the following:

\begin{equation}
\begin{array}{l}
f^2 = \dac{4(2D-3)(D+2)(D^2-9)c_0}{3D^2(D-2)c_1}.
\end{array} \label{pert3}
\end{equation}

The above dependence of the frequency on the number of extra dimensions clearly shows a WKB-eikonal pattern. Namely, the frequency grows with the number of extra dimensions (as one would expect
in a eikonal scheme). Thus, the larger is $D$, the closer we are to a genuine eikonal behavior.

\section{Influence of matter}

If we substitute (\ref{b0}) and the subregime condition (\ref{coup}) into (\ref{friedmann1}) and we choose $c_1 = 1$ as normalization, we obtain the following expressions:

\begin{equation}
\begin{array}{l}
G_{N, eff} = \dac{2(2D-3)(D+1)}{(D-2)(D-3)},~~\Lambda_{4D, eff} = 0.
\end{array} \label{friedmann2}
\end{equation}

Obviously this regime is trivial, but if we add a matter in form of perfect fluid with linear equation of state $P=\omega \rho$ , we can hope to recover regular $(3+1)$-dimensional Friedmann
regime. So we numerically integrate equations of motion which are (\ref{eq_my_1})--(\ref{eq_my_3}) considering the subregime (\ref{coup}) introduced in the previous section. Also we added
continuity equation for matter in the form

\begin{equation}
\begin{array}{l}
\dot\rho(t) + \(3H(t) + D\dac{\dot b(t)}{b(t)}\)\(1+\omega\)\rho(t) = 0.
\end{array} \label{cont_eq}
\end{equation}

One can clearly see that for regime with $b(t) = b_0 \equiv \const$ the second term in first brackets in nullified and so standard $(3+1)$-dimensional density scaling is recovered.
Also, Friedmann constraint is also modified in presence of matter:

\begin{equation}
\begin{array}{l}
G_{N, eff} H(t)^2 = \Lambda_{4D, eff} + \rho(t).
\end{array} \label{friedmann2}
\end{equation}

\begin{figure}
\includegraphics[width=1.0\textwidth, angle=0]{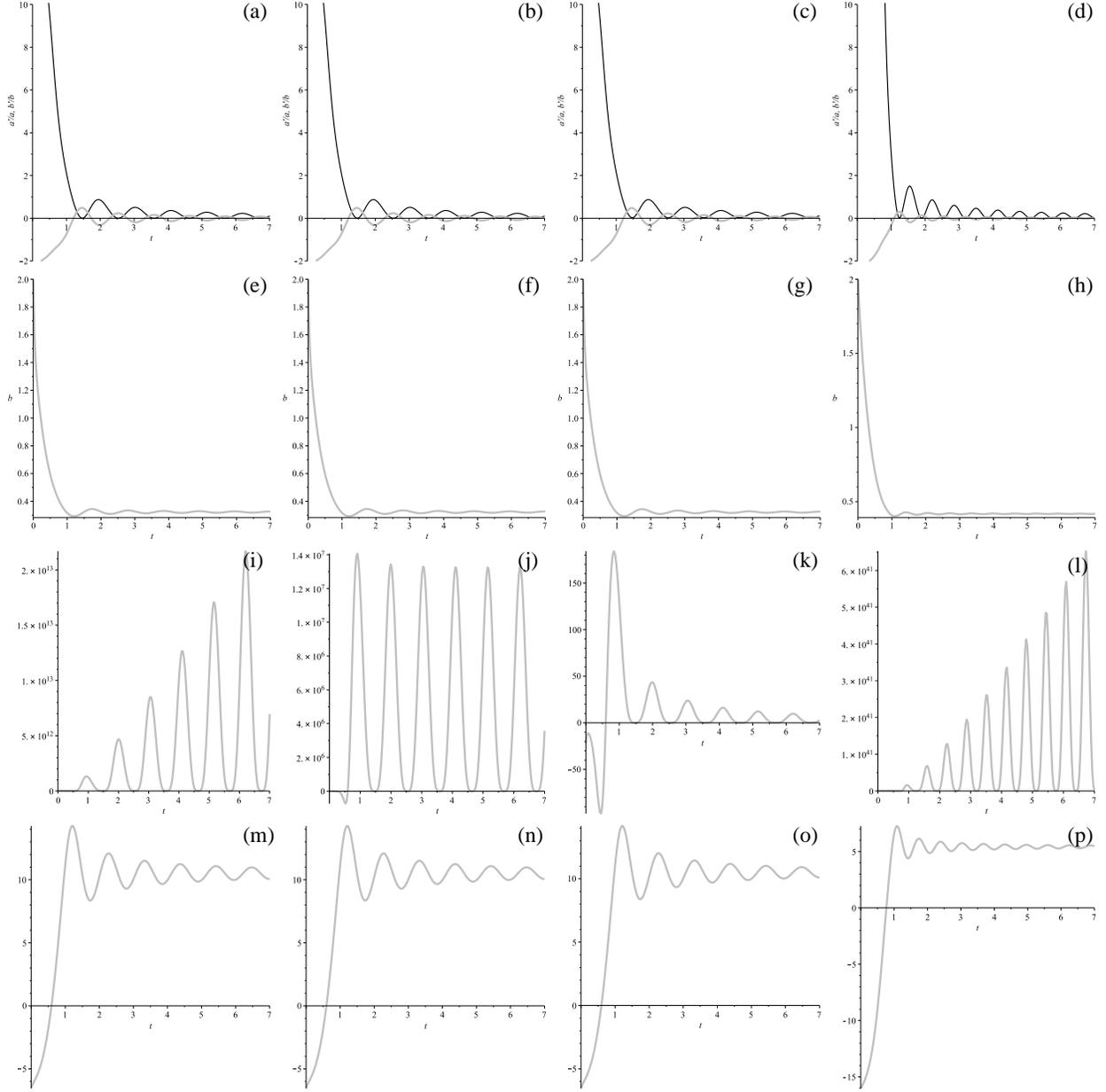}
\caption{Behavior of Hubble parameters, size of extra dimensions, density-to-squared Hubble ratio and effective Newtonian constant with time for our model (see text for details).}
\label{fig1matter}
\end{figure}

As we said,
we solved numerically system under consideration to see the future and past behavior. The results are presented in Fig.~\ref{fig1matter}. There the first row ((a), (b), (c) and (d) panels)
presents the Hubble parameters associated with $(3+1)$ (in black) and extra dimensions (in grey), the second row ((e), (f), (g) and (h) panels) shows us the size of extra dimensions $b(t)$,
the third row ((i), (j), (k) and (l) panels) visualizes $G_{N, eff}H(t)^2/\rho(t)$ (we comment on it later) while fourth ((m), (n), (o) and (p) panels) row demonstrate effective Newtonian
constant $G_{N, eff}$ from (\ref{friedmann2}). All rows have four columns which correspond to four cases -- first column ((a), (e), (i) and (m) panels) corresponds to $D=6$, $\omega = 0.6$ case,
second column ((b), (f), (j) and (n) panels) corresponds to $D=6$, $\omega = 0$ case, third column ((c), (g), (k) and (o) panels) corresponds to $D=6$, $\omega = - 0.6$ case while the last column
((d), (h), (l) and  (p) panels) corresponds to $D=16$, $\omega=0.6$ case.
This way if we compare the results between first, second and third columns, we can derive the influence of the equation of state on the
dynamics while comparison of the first and fourth gives us effect of the number of extra dimensions.

Then, by comparing first three columns we can decide that the effect of the equation of state demonstrates only in the $G_{N, eff}H(t)^2/\rho(t)$ ratio (third row, (i), (j), (k) and (l) panels).
We can see that the influence of matter on the dynamics is diminished for $\omega > 0$ and significant for $\omega < 0$. We comment more about it in Discussions section.
In contrary
the effect of the number of extra dimensions is visible for all cases -- first of all, the period of oscillations is decreased for all variables (Hubble parameters, size of extra dimensions,
etc) while its amplitude seems to decrease for all cases except Hubble parameters, where it increases instead.

\begin{figure}
\includegraphics[width=0.92\textwidth, angle=0]{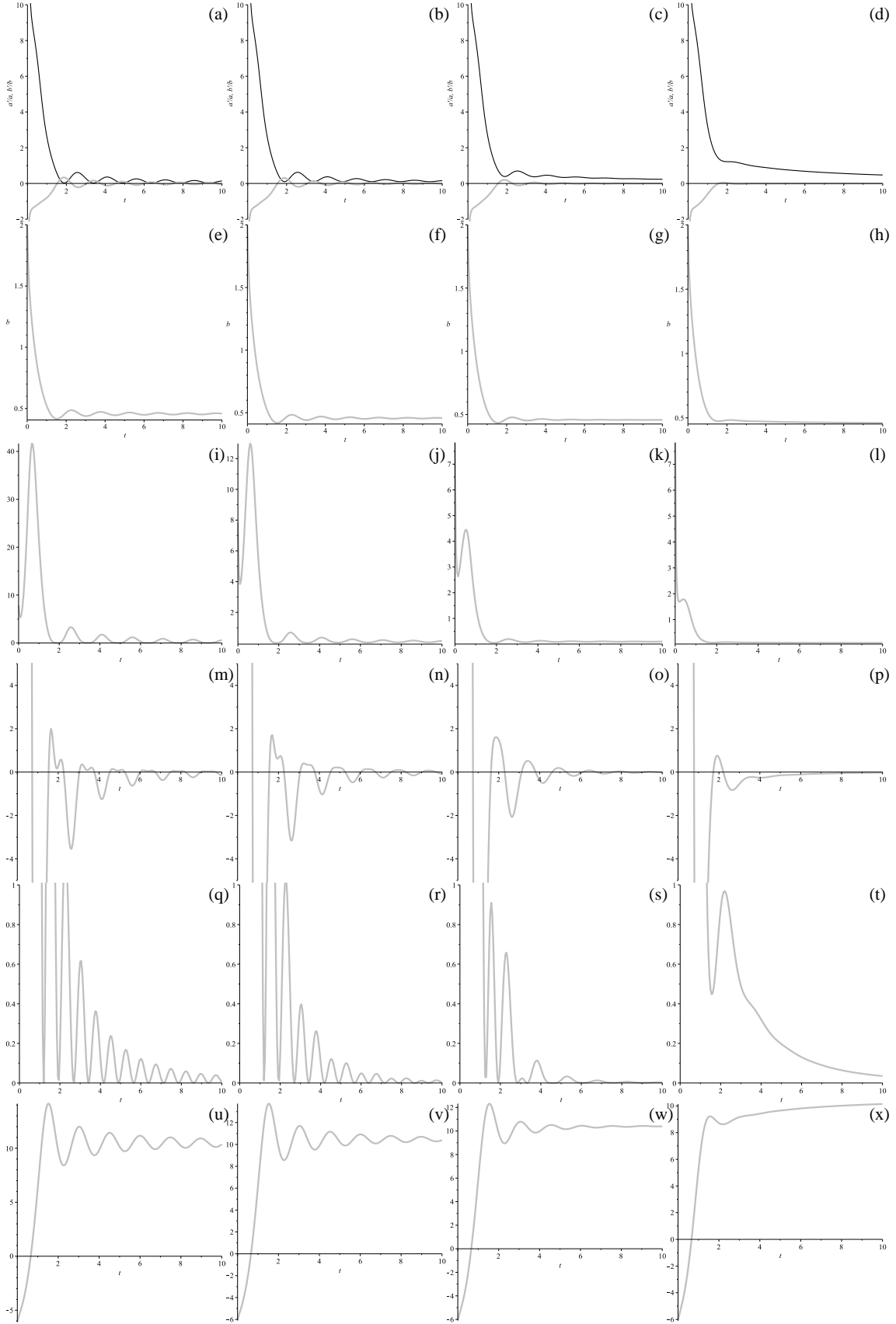}
\caption{Similar as Fig.~\ref{fig1matter} but in the vicinity of $\omega \gtrsim -1$ (see text for details).}
\label{fig2matter}
\end{figure}

Also at that point it is necessary to note that $b(t)$ is reaching its predicted value (\ref{b0}) and the period of oscillations is in agreement with (\ref{pert3}). This way we ensure that
our model is reached indeed in our numerical study. Similar to the general compactification from geometric frustration regime~\cite{CGP1, CGP2}, this regime is also achieved on all
initial conditions which lead to the former regime with an additional (\ref{coup}) relation between couplings taken into account.

Regarding the influence of the equation of state, there is one more important point -- for strongly negative equations of state ($\omega \lesssim -0.8$) we detected qualitative change in
behavior of main variables. This change is quantitatively demonstrated in Fig.~\ref{fig2matter}. In there we presented Hubble parameters (first row, (a)--(d) panels), size of extra dimensions
$b(t)$ (second row, (e)--(h) panels), $H(t)^2/\rho(t)$ value (third row, (i)--(l) panels), Friedmann constraint (\ref{friedmann2}) (fourth row, (m)--(p) panels), effective $\Lambda$-term
(fifth row, (q)--(t) panels) and effective Newtonian constant (sixth row, (u)--(x) panels). These parameters are presented for four consecutive values for the equation of state:
$\omega = - 0.6$ (first column, (a), (e), (i), (m), (q) and (u) panels), $\omega = - 0.7$ (second column, (b), (f), (j), (n), (r) and (v) panels), $\omega = - 0.8$ (third column, (c), (g), (k),
(o), (s) and (w) panels) and $\omega = - 0.9$ (fourth column, (d), (h), (l), (p), (t) and (x) panels). One can see that the first column resemble typical behavior of Fig.~\ref{fig1matter} but
with further decrease of the equation of state the behavior differ more and more. It looks like the oscillations, widely presented in Fig.~\ref{fig1matter}, decaying for strongly negative
equations of state in Fig.~\ref{fig2matter}. If we can make mechanical analogy, the friction in Fig.~\ref{fig2matter} is so high that it prevents oscillations from occurring and we have
analogue of ``slow-roll'' instead.



The analysis of the previous section explains oscillatory behavior, but from numerical investigation (see Figs~\ref{fig1matter} and \ref{fig2matter}) we can see that the oscillations are damped. The explanation
for this lies in the fact that we perturb equations around exact solution $\dot H = H = 0$, $\ddot b = \dot b = 0$ plus (\ref{b0}) and (\ref{coup}). But in reality this asymptotic regime
is achieved pretty slowly so that instead of (\ref{pert2}) we should use full system which is very complicated for exact analytical analysis but has exponential damping, which is clearly seen
from Figs.~\ref{fig1matter} and \ref{fig2matter}.

Let us note an interesting thing -- formally we do not bound amplitude of described oscillations, but from Figs.~\ref{fig1matter} and \ref{fig2matter} one can see that $H(t)$ oscillate in a
way that $H(t) \geqslant 0$ always. It is easy to see why is that -- if one substitute $H(t) = 0$ into (\ref{eq_my_1}) with (\ref{coup}) taken into account, the resulting equation links
$b(t)^2$ with $\dot b(t)^2$. If we further substitute this resulting relation into (\ref{eq_my_3}) with (\ref{coup}) and $H(t)=0$, we arrive to $\dot H = 0$. This way one can see that $H(t)$
cannot cross zero and since we put $H(t) \geqslant 0$ from the beginning it always remains non-negative.

\section{Discussion and conclusions}

In this paper we investigated the dynamical compactification  in Einstein-Gauss-Bonnet gravity with arbitrary number of extra dimensions in the presence of a perfect fluid matter source. A
detailed analysis was performed for the case where the theory does not admit a maximally symmetric solution. For a manifold which is a warped product of a four dimensional FRW manifold times a
constant (negative) curvature compact Euclidean manifold it was found that  there exists a subregime which leads to $H_a \to 0$  and $b(t) \to \const$ via damped oscillations and so  standard
$(3+1)$-dimensional Friedmann cosmology with perfect fluid is asymptotically recovered at  least at the level of the background equations.

This regime appears to be sensitive to the equation of state of the matter , and the situation here is similar to the GR limit in the
$R+R^2$ gravity. In the latter theory there are two additional degrees of freedom connected with higher derivatives of the scale
factor, so they (in contrast to Gauss-Bonnet gravity) are present in the isotropic (3+1) dimension case. They oscillate harmonically in
the low-curvature limit
(this is the feature of quadratic gravity, for higher order corrections the oscillation are anharmonic, see  ~\cite{BIT}).
and can be expressed in the form of a massive scalar field. The effective equation of state for such field
during oscillation is known to be $w=0$, so any matter with negative pressure would decay less rapidly, and, as a result,
dominates for late time. On the contrary, a matter with a positive pressure should be subdominant in the low-curvature regime,
and universe becomes dominated by gravitational degrees of freedom.

In our model the situation is quite similar. We have to additional degrees of freedom -- the scale factor of ``inner'' dimensions and its
derivative. They oscillate harmonically in the first approximation. Our numerical results show that matter content of the Universe
become dominant for $w<0$ while for $w>0$ influence of the ``inner'' degrees of freedom may be important at later time since it decays
slowly than the matter.

Another interesting numerical  effect of the equation of state is the fact that for $\omega \lesssim -0.8$ an oscillatory behavior, presented in
Fig.~\ref{fig1matter}, is replaced with monotonic one, presented in Fig.~\ref{fig2matter}. Change in the behavior could be treated with the following mechanical interpretation -- the
friction is
increased and through this oscillations which are clearly seen in Fig.~\ref{fig1matter} are almost decayed by the last column of Fig.~\ref{fig2matter}.

As for the influence of the number of extra dimensions, it manifests itself in the growing of frequency of oscillations for fixed coupling
constants, which is also clear from the analytical estimation. From the technical point of view, the field equations have been written down in order to display explicitly the dependence on the
number of extra dimensions. The reason is that many of the relevant physical quantities analyzed here depend on the number of extra dimensions in the same way as one would expect from a large
$N$ expansion in gauge theory (in which $D$ plays the role of $N$). In particular, a clear ``eikonal-like'' pattern emerges. Thus, the present results strongly suggest the importance to develop
a systematic $1/D$ expansion in EGB cosmology. We hope to come back on this interesting issue in a future publication.

To conclude, the aims of this paper were reached -- we described new regime which is part of previously described model which allows natural and viable dynamical compactification. We
demonstrated that our described regime has viable Friedmann asymptotic behavior so that at late time it cannot be distinguished from ``standard'' (3+1)-dimensional Friedmann model.
Additionally, effective $\Lambda$-term for this model has geometrical nature which could potentially solve ``cosmological constant problem'' without involving any additional physics, though 
suffering from the fine tuning problem -- the equation (\ref{coup}) which is exact for zero four-dimensional $\Lambda_{4D,eff}$ should be almost exactly satisfied.

\begin{acknowledgments}
A.G. was supported by FONDECYT grant No 1150246. F.C. was supported by FONDECYT grant No 1160137. The Centro de Estudios Cientificos (CECs) is funded by the Chilean Government through the
Centers of Excellence Base Financing Program of Conicyt. The work of S.A.P. was supported by FAPEMA.
  The work of A.T. is supported by RFBR grant 14-02-00894, and 
  partially supported by the Russian Government Program of Competitive Growth of Kazan Federal University. A.T. thanks Universidad Austral
de Chile (Valdivia). where this work was initiated, for hospitality. 
\end{acknowledgments}

\end{document}